\documentclass[a4paper]{jpconf}
\usepackage{graphicx}
\usepackage{amsmath,amssymb,amsfonts,float}
\newcommand{\bolk}{\mathbf{k}}
\newcommand{\bolq}{\mathbf{q}}
\newcommand{\bole}{\mathbf{e}}
\newcommand{\bolQ}{\mathbf{Q}}

\newcommand{\bolr}{\mathbf{r}}

\newcommand{\bolK}{\mathbf{K}}
\newcommand{\bolp}{\mathbf{p}}

\newcommand{\VEV}[1]{\langle #1 \rangle}  
\newcommand{\be}{\text{e}}

\newsavebox{\dotdot}
\savebox{\dotdot}[3mm]{\shortstack{\circle*{0.8}\\ \\ \circle*{0.8}}}

\begin{document}
\title{Dilute-Bose-Gas Approach to ground state phases 
of 3D quantum helimagnets under high magnetic field}

\author{Hiroaki T. Ueda$^1$, Keisuke Totsuka$^1$, Tsutomu Momoi$^2$}

\address{$^1$Yukawa Institute for Theoretical Physics, Kyoto University, Kitashirakawa
Oiwake-Cho, Kyoto 606-8502, Japan\\
$^2$Condensed Matter Theory Laboratory, RIKEN, Wako, Saitama 351-0198, Japan}

\begin{abstract}
We study high-field phase diagram and low-energy excitations of three-dimensional quantum helimagnets. Slightly below the saturation field, the emergence of magnetic order may be mathematically viewed as Bose-Einstein condensation (BEC) of magnons. The method of dilute Bose gas enables an unbiased quantitative analysis of quantum effects in three-dimensional helimagnets and thereby three phases are found: cone, coplanar fan and an attraction-dominant one. To investigate the last phase, we extend the usual BEC approach so that we can handle 2-magnon bound states. In the case of 2-magnon BEC, the transverse magnetization vanishes and long-range order occurs in the quadrupolar channel (spin-nematic phase). As an application, we map out the phase diagram of a 3D helimagnet which consists of frustrated $J_1$-$J_2$ chains coupled by an interchain interaction $J_3$.
\end{abstract}

\begin{figure}[h]
\begin{minipage}{18pc}
\includegraphics[width=18pc]{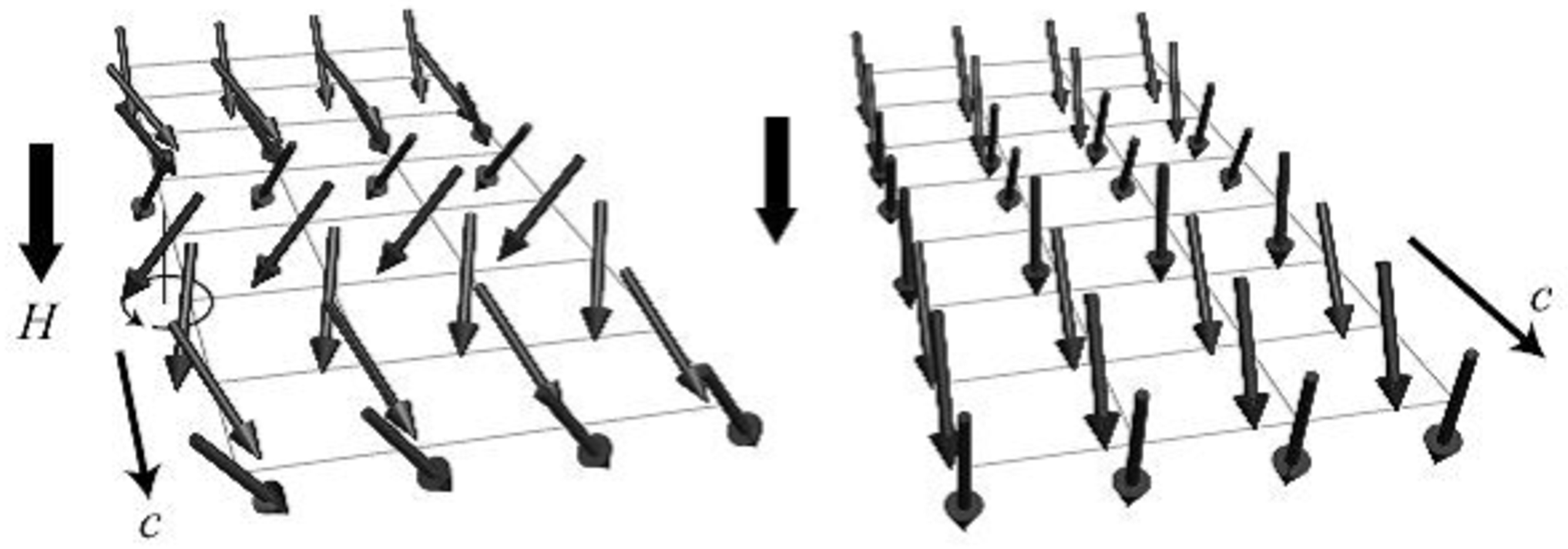}
\caption{\label{fig:cone-fan}Two spin structures considered here: `cone'
(left) and `fan' (right). In the fan structure, spins are lying in a
single plane. The Q-vector is pointing along the c-axis.}
\end{minipage}\hspace{2pc}%
\begin{minipage}{17pc}
\includegraphics[width=17pc]{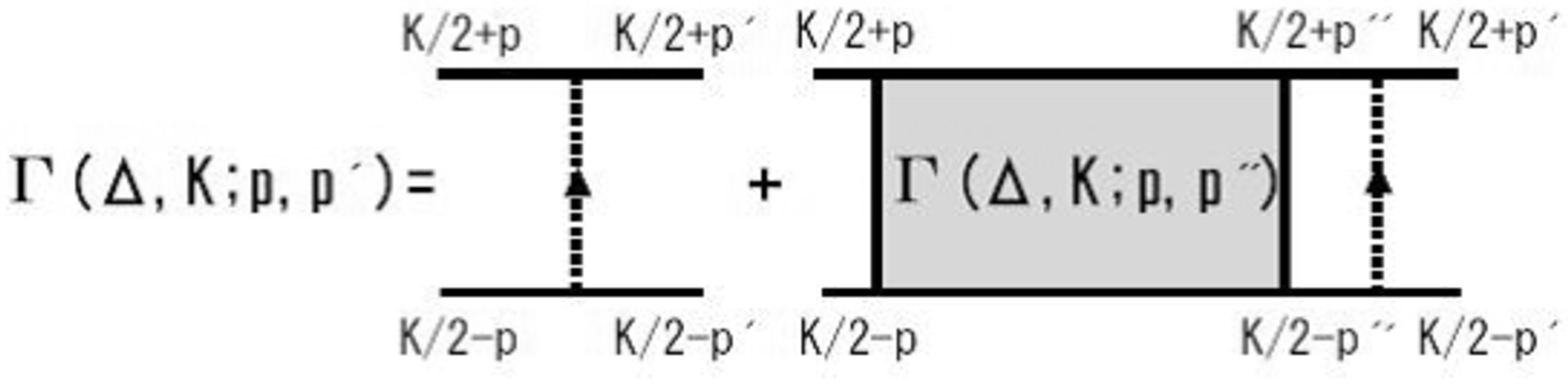}
\caption{\label{fig:ladder}Ladder diagram of the interaction vertex}
\end{minipage} 
\end{figure}

Magnetic frustration introduces several competing states which are 
energetically close to each other and thereby destabilizes 
simple ordered states.  
One way to compromise two or more competing orders is to 
assume a helical (spiral) spin structure\cite{Yoshimori}. 
In this letter, we discuss the high-field behavior of 
a spin-1/2 Heisenberg model with generic interactions: 
$
H=\sum_{\langle i,j\rangle}J_{ij}\,{\bf S}_i{\cdot}{\bf S}_j 
+ \text{H}\sum_{j}S^{z}_{j} .
$
For the simplest case with one magnetic ion per unit cell, one can 
easily find the classical ground state by minimizing the Fourier 
transform of the exchange interactions:
\begin{equation}
\epsilon(\bolq)=\sum_{j} \frac{1}{2} J_{ij}\cos \left(
\bolq {\cdot}(\bolr_i-\bolr_j)\right)\ ,
\end{equation}
where the summation is taken over all $j$-sites connected to 
the $i$-site by $J_{ij}$.    
When $\epsilon(\bolq)$ takes its minima $\epsilon_{\text{min}}$ 
at $\bolq=\pm \bolQ$, helical order with the wave number $\bolQ$ 
or $-\bolQ$ appears 
($\pm \text{\bf Q}$ are not equivalent to each other).  
When the external magnetic field is applied in a direction perpendicular 
to the spiral plane, the spiral at H=0 is smoothly deformed 
into the so-called {\em cone} state (figure \ref{fig:cone-fan}) 
as H is increased 
and this persists until all spins eventually get polarized 
at the saturation field $\text{H}_{\text{c}}$ \cite{Nagamiya}.
%When the system has an easy-plane anisotropy and 
%the external field is applied in the 
%spiral plane, on the other hand, the system undergoes a (first-order) 
%metamagnetic transition into a coplanar {\em fan} 
%phase\cite{Nagamiya}. 

One of the simplest models which exhibit, at least in the classical 
limit, the helical order is a three-dimensionally coupled 
Heisenberg chains with nearest-neighbor- 
(NN) $J_1$ and next-nearest-neighbor (NNN) $J_2$ coupling. 
Rise of multiferroics revives study of helimagnetism and  
many compounds which contain these 1D-chains 
as subsystems have been reported 
(see, for instance, TABLE I. in reference \cite{hase}). 
For example, a helimagnetic
material LiCuVO$_4$ may be viewed as coupled quantum
S = 1/2 $J_1$-$J_2$ chains and 
exhibits helical spin order, which is expected from the classical 
theories, and ferroelectricity simultaneously under moderate magnetic field. 
When the field is very high, on the other hand, this compound shows modulated
collinear order, which contradicts with the aforementioned
classical prediction, and this may suggest that quantum
fluctuation plays an important role \cite{LiCuVO4}. 
Therefore, it would be interesting to explore the possibility 
that quantum fluctuation replaces the classical cone state with 
other stable ones. 

By using dilute-Bose-gas approach, Batyev and Braginskii \cite{Batyev-1} 
investigated magnetic structures near saturation (H=H$_{\text{c}}$)   
and concluded that a new coplanar fan phase appears if a certain
condition for the bosonic interactions is satisfied. 
Our aim in this letter is to determine the stable spin configurations 
of a specific 3D spin-1/2 helimagnet in a fully quantum-mechanical manner.  

%%%%%%%%%%%%%%%%%%%%%%%%%%%%%%%%%%%%%%%%%%%%%%%%%%%%%%%%%%%%%%%%%%%%
%%%%%%%%%%%%%%%%%%%%%%%%%%%%%%%%%%%%%%%%%%%%%%%%%%%%%%%%%%%%%%%%%
{\em General Formalism}-- 
%%%%%%%%%%%%%%%%%%%%%%%%%%%%%%%%%%%%%%%%%%%%%%%%%%%%%%%%%%%%%%%%%
By taking fully polarized state as the vacuum and treating spin-flips 
as hardcore bosons 
$S^z_l=-1/2+\beta^\dagger_l\beta_l$, 
$\ S_l^+ =\beta_l^\dagger,\ S_l^- =\beta_l$, 
we can rewrite the spin Hamiltonian as:
\begin{equation}
H = \sum_{q}(\omega (\bolq) - \mu)\beta^\dagger_{\bolq}\beta_{\bolq}
+\frac{1}{2N}  \sum_{\bolq,\bolk,\bolk^\prime}  V_{\bolq} 
\beta_{\bolk+\bolq}^\dagger\beta_{\bolk^\prime-\bolq}^\dagger
\beta_{\bolk}\beta_{\bolk^\prime},
\end{equation}
\begin{equation}
\begin{split}
\omega(\bolq) =\epsilon(\bolq)-\epsilon_{\text{min}}\ ,\ \ 
\mu={\rm H}_{\text{c}}-{\rm H}\ ,\ \ 
{\rm H}_{\text{c}}=\epsilon({\bf 0})-\epsilon_{\text{min}}\ ,\ \ 
V(\bolq) =2(\epsilon(\bolq)+U)\ ,
\end{split}
\end{equation}
where $U({\rightarrow}\infty)$ has been added to insure the hard-core 
constraint and the minimum of $\epsilon({\bf q})$ is taken 
$\epsilon_{\text{min}}=\epsilon(\pm{\bf Q})$ for helimagnets. 
In what follows, we consider a cubic lattice 
(we reserve $(a,b,c)$ to label the three crystal axes)
and assume that helical- and ferromagnetic/antiferromagnetic 
order occur along the $c$-axis and in the $ab$ plane, respectively 
(i.e. $\mathbf{Q}=(0,0,Q)$ or $\mathbf{Q}=(\pi,\pi,Q)$). 
On general grounds, we may expect that magnon BEC occurs 
when the external field is $\text{H} < \text{H}_{\text{c}}$ ($\mu>0$).  
%%%%%%%%%%%%%%%%

The thermal potential per site $E/N$ of the dilute Bose gas 
is determined by the renormalized interaction ($t$-matrix) among 
the condensed bosons at ${\bf q}=\pm {\bf Q}$ 
and the ground state densities $\rho_{\pm {\bf Q}}$ are obtained 
by minimizing $E/N$. 
If we denote the renormalized interactions between 
the like bosons and that between the different ones respectively 
as $\Gamma_1$ and $\Gamma_2$, the energy density $E/N$ is given by
\begin{equation}
\frac{E}{N} =\frac{1}{2}\Gamma_1
\left(\rho_{\bolQ}^2+\rho_{-\bolQ}^2\right)
+\Gamma_2 \, \rho_{\bolQ}\, \rho_{-\bolQ} 
- \mu(\rho_{\bolQ}+\rho_{-\bolQ}),
\label{EffPotential}
\end{equation}
where $\rho_{\bolq}=|\VEV{\beta_{\bolq}}|^2/N$.
Different phases appear according to the values of $\Gamma_{1,2}$.  
When $\Gamma_2 > \Gamma_1 >0$, $E/N$ is minimal for the choice 
$\rho_{\bolQ} =\rho=\mu/\Gamma_1$,  
$\rho_{-\bolQ}=0$ (or vice versa) and 
$E/N=- \mu^2/(2\Gamma_1)$.
Then, the spin configuration is determined as:
\begin{equation}
\begin{split}
& \VEV{\beta_l}=\sqrt{\rho}\exp\{\pm i({\bf Q}{\cdot}{\bf R}_l + \theta)\}
\ ,\ \ \VEV{S_l^z}=-\frac{1}{2}+\rho\ ,\\
& \VEV{S_l^x} = \sqrt{\rho}\cos ({\bf Q}{\cdot}{\bf R}_l + \theta)\ ,\ \ 
\VEV{S_l^y} = \mp\sqrt{\rho}\sin ({\bf Q}{\cdot}{\bf R}_l + \theta).
\end{split}
\end{equation}
That is, the cone state (the left panel of figure \ref{fig:cone-fan}), 
which exists already in the classical case \cite{Nagamiya},  
is favored for $\Gamma_{2}>\Gamma_{1}$. 

If $\Gamma_1>\Gamma_2$ and $\Gamma\equiv \Gamma_1+\Gamma_2>0$, 
on the other hand, 
the two modes condense simultaneously and 
the ground state is determined as:
$\rho_{\bolQ} =\rho_{-\bolQ}=\rho^\prime=\mu/\Gamma$, 
$\frac{E}{N}=- \mu^2/\Gamma$ 
\begin{equation}
\begin{split}
\VEV{\beta_l}& =\sqrt{\rho^\prime}
\left\{ 
\be^{i({\bf Q}{\cdot}{\bf R}_l+\theta_1)}
+\be^{i(-{\bf Q}{\cdot}{\bf R}_l+\theta_2)}
\right\}  \ ,\\
\VEV{S_l^z}
 &=-\frac{1}{2}+4\rho^\prime\cos^2 ({\bf Q}{\cdot} {\bf R}_l
+\frac{\theta_1-\theta_2}{2})\ ,\ \ 
\VEV{S_{l}^{\pm}}
 =2\sqrt{\rho^\prime}\cos ({\bf Q}{\cdot} {\bf R}_l
+\frac{\theta_1-\theta_2}{2})\be^{\mp i\frac{\theta_1+\theta_2}{2}}\ .
\label{superfan}
\end{split}
\end{equation}
The existence of the two phases $\theta_{1}$ and $\theta_{2}$, 
which correspond respectively to  
the two condensates $\VEV{\beta_{\bf Q}}$ and $\VEV{\beta_{-{\bf Q}}}$, 
lead to two different low-energy excitations. 
Since ${\VEV{S_l^y}}/{\VEV{S_l^x}}=-\tan\frac{\theta_1+\theta_2}{2}$, 
the spins assume a coplanar configuration ({\em fan}) shown 
in the right panel of figure \ref{fig:cone-fan}.  
%%%%%%%%%%%%%%%%%%%%
According to the standard theory \cite{MF-review}, 
this phase does not exhibit ferroelectricity. 

In these two Bose condensed phases, there exists two low-energy modes at $\bolq = \pm\bolQ$ and the low-energy physics is described by the effective Lagrangian
with U(1)$\times$ U(1) symmetry; one comes from the axial
(around the external field) symmetry and the other from an
emergent translational symmetry. In the cone phase, only one
of the two bosons condenses and there is one gapless Goldstone
mode (the two U(1)s are no longer independent). Meanwhile,
the fan phase breaks both symmetries and has two types
of gapless Goldstone modes(for more detail, see reference \cite{UT}).

When $\Gamma_1<0$ or $\Gamma_1+\Gamma_2<0$, 
low-energy bosons around $\mathbf{q}=\pm \mathbf{Q}$ 
attract each other. 
If the energy (\ref{EffPotential}) is taken literally, 
one may expect, on general grounds, first order transitions to occur 
(provided that the cubic terms of $\rho$ are positive). 
%In some cases, this scenario may be the case. 
However, eq.(\ref{EffPotential}) is based on the assumption 
that magnon BEC occurs in the {\em single-magnon channel} and 
may not work when we expect magnon bound states stabilized by 
strong attraction.  
%In fact, this conditions for $\Gamma_{1,2}$ implies nothing but 
%instability in the one-magnon condensates. 
In fact, according to the standard Bethe-Salpeter method, 
divergingly large (renormalized) attraction implies the existence of 
stable 2-magnon bound states as a pole in the two-particle Green's 
function corresponds to a bound state.  
Therefore, the simple single-magnon BECs will give way to those in 
multi-particle channel. 
To check the above scenario, we solve the 2-magnon scattering problem. 
We denote the ladder diagram as
$\Gamma(\Delta,\bolK;\bolp,\bolp^\prime)$ 
(see figure \ref{fig:ladder}), where we take the total energy of initial state 
as $-2\mu-\Delta$. 
Then, on the fully saturated ground state, the exact scattering amplitude 
$M(\Delta,\bolK;\bolp,\bolp^\prime)=\Gamma(\Delta,\bolK;\bolp,\bolp^\prime)
+\Gamma(\Delta,\bolK;\bolp,-\bolp^\prime)$ is obtained by solving 
the following integral equation:
\begin{equation}
M(\Delta,\bolK;\bolp,\bolp^\prime)=V(\bolp^\prime-\bolp)+V(-\bolp^\prime-\bolp)-\frac{1}{2N}\sum_{\bolp^{\prime\prime}}\frac{M(\Delta,\bolK;\bolp,\bolp^{\prime\prime})(V(\bolp^\prime-\bolp^{\prime\prime})+V(-\bolp^\prime-\bolp^{\prime\prime}))}{\omega(\bolk/2+\bolp^{\prime\prime})+\omega(\bolk/2-\bolp^{\prime\prime})+\Delta-i0^+}\ .
\label{laddereq}
\end{equation}
Even in the presence of the boson condensate, as long as it is dilute, 
we can safely use this scattering amplitude. 
Hence, one obtains the renormalized interaction $\Gamma_{1}$ and $\Gamma_{2}$  
as $\Gamma_1 =(1/2)M(0,2\bolQ;0,0),\ \Gamma_2 =M(0,0;\bolQ,\bolQ)$  \cite{Nikuni-Shiba-1}. 
Moreover, the pole of $M(\Delta,\bolK;\bolp,\bolp^\prime)$ gives the stable bound state. 

\begin{figure}[h]
\includegraphics[width=18pc]{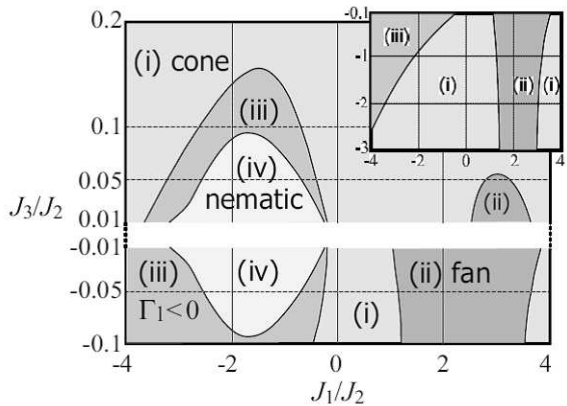}\hspace{2pc}%
\begin{minipage}[b]{18pc}\caption{Phase diagram slightly below saturation 
($\text{H}\lesssim \text{H}_{\text{c}}$) mapped out 
in $(J_{1},J_{3})$-plane 
($J_{2}(>0)$ is used to set the energy unit). 
(i) {\em cone phase} and (ii) {\em coplanar fan phase} are 
one-magnon condensed phases. 
In the phase-(iii), attractive interaction may imply instabilities 
toward other phases e.g. conventional ferromagnetic one or 
more exotic multipolar ones. 
The phase (iv) ({\em nematic}) is characterized by the condensation of the 2-magnon
bound state and leads to the nematic order in the transverse direction.  
The region $|J_{3}|/J_{1}\ll 1$ is omitted for a reason described in the text. 
Inset: The same phase diagram for the large negative 
interchain coupling ($-J_{3}>0.1$). %
\label{Fig:quasi1D}}
\end{minipage}
\end{figure}

%%%%%%%%%%%%%%%%%%%%%%%%%%%%%%%%%%%%%%%%%%%%%%%%%%%%%%%%%%%
{\em Coupled $J_1$-$J_2$ model}-- 
%%%%%%%%%%%%%%%%%%%%%%%%%%%%%%%%%%%%%%%%%%%%%%%%%%%%%%%%%%%
Having established the formalism, we proceed to investigating 
a specific model--a frustrated spin-1/2 model on a simple cubic lattice 
whose Hamiltonian is given by
\begin{equation}
H=
\sum_{\bolr,i=a,b} \left\{J_1 {\bf S}_{\bolr}{\cdot}
{\bf S}_{\bolr+\hat{\bole}_{c}}
+J_{2}{\bf S}_{\bolr}{\cdot}{\bf S}_{\bolr+2\hat{\bole}_{c}}
+J_3{\bf S}_{\bolr}{\cdot}{\bf S}_{\bolr+\hat{\bole}_{i}}\right\}+ \text{H}\sum_{j}S^{z}_{j} ,
\label{eqn:J1-J2-J3}
\end{equation}
In (\ref{eqn:J1-J2-J3}), 
the $J_{1}$-$J_{2}$ chains are running in the $c$-direction 
and $J_{3}$ controls the coupling among adjacent chains. 
The wave number ${\bf Q}$ characterizing the condensate is given 
either by $\bolQ=(0,0,Q)$ ($J_{3}<0$) or 
by $\bolQ=(\pi,\pi,Q)$ ($J_{3}>0$) where 
$Q=\arccos (-{J_1}/{4J_2})$.  Hence the spiral occurs in the
$c$-direction. 
To determine the spin structure of our $J_{1}$-$J_{2}$-$J_{3}$ model, 
we solved eq.(\ref{laddereq}) by assuming the following 
form \cite{Batyev-1}:  
$M(\Delta,\bolK;\bolp,\bolp^\prime)=\VEV{M}+A_1\cos p_c^\prime
+A_2\cos 2p_c^\prime+A_3\cos p_a^\prime+A_4\cos p_b^\prime$, 
where $\VEV{M}$ and $A_i$ are functions of $\Delta,\bolK$ and $\bolp$.  
After some numerical calculations, we obtained the phase diagram  
shown in figure \ref{Fig:quasi1D}. We show only the frustrated region 
$-4 \le J_{1}/J_{2} \le 4$, where cone structure with 
incommensurate ${\bf Q}$ is expected classically.  
%%%%%%%%%%%%%%%%%%%%%%%%%%%%%%%%%%%%%%%%%%%%%%%%%%%%%%%%%
%%%%%%%%%%%%%%%%%%%%%%%%%%%%%%%%%%%%%%%%%%%%%%%%%%%%%%%

For $J_3 \rightarrow 0$, low-energy quantum fluctuation 
destabilizes $\Gamma$  
and our approach cannot be extended to $J_3=0$ 
continuously ($\Gamma$ becomes $O(J_3^{1/2}$) and 
$\Gamma_1 \rightarrow \Gamma_2$ at the leading order in $J_3$).

Finally, we compare the phase diagram shown in figure \ref{Fig:quasi1D} 
with that of the 1D $J_{1}$-$J_{2}$ chain ($J_{3}=0$).  Let us begin with 
the case $J_{1}>0$.  Near saturation, two dominant phases are found 
in 1D \cite{1D-J1-J2}: 
(i) `chiral phase (VC)' with finite vector chirality parallel 
to the magnetic field and (ii) `TL2' phase where the system is 
described by two Tomonaga-Luttinger (TL) liquids.  
Obviously, the former turns, after switching on $J_{3}$, into 
the cone phase. A close inspection of the two gapless TL modes 
near saturation tells us that the TL2 phase should be identified 
with the fan phase here. 

For the ferromagnetic case $J_1 < 0$, 
BECs of $n$-bound magnon states ($n\geq$ 2) are expected in 1D chain \cite{bound}. 
We found how strong inter-chain coupling destabilizes the 2-bound magnon BEC. 
The study of the stability of higher $n$-bound magnon state ($n\geq 3$) 
in 3D is a future problem.

\end{document}